# Multipoint Monitoring of Instantaneous Amplitude, Frequency, Phase and Sequence of Vibrations Using Concatenated Modal Interferometers

**Kalipada Chatterjee[1], Venugopal Arumuru[2], Dhananjay Patil[2], and Rajan Jha[1*]**

[1]*Nanophotonics and Plasmonics Laboratory, School of Basic Sciences, Indian Institute of Technology Bhubaneswar, 752050, India*

[2]*Applied Fluids Group, School of Mechanical Sciences, Indian Institute of Technology Bhubaneswar, 752050, India*

*rjha@iitbbs.ac.in

Abstract

Concatenated modal interferometers based multipoint sensing system for detection of instantaneous amplitude, frequency, and phase of mechanical vibrations is proposed and demonstrated. The sensor probes are fabricated using identical photonic crystal fiber (PCF) sections and integrated along a single fiber channel to act as a compact and efficient sensing system. Individual probes operate independently producing a resultant signal that is a superposition of each interferometer response signal. By analyzing the resultant signals, information about the measurand field at each location is realized. Such a sensing system would find wide applications at industrial, infrastructural, and medical fronts for monitoring various unsteady physical phenomena.

Real-time monitoring of mechanical vibrations over multiple locations enables structural health monitoring[1], posture recognition[2], and industrial surveillance. Such mechanical vibrations are characterized by their physical parameters viz. frequency, amplitude, and relative phase that differ over a broad range depending on the measurand field's functionality. Detection of the aforementioned vibration parameters has been prominently achieved by conventional electronic vibration sensors at many fronts. However, the performance of these devices is influenced by power fluctuations, environmental temperature variations, and stray electromagnetic field interference. Alternatively, optical fiber-based vibration sensors have proven to be a potential class of devices that function on the principle of modulation of intensity, wavelength composition, or phase of the propagating light in the presence of an external measurand field. These sensors are compact, accurate, reconfigurable, corrosion-resistant, temperature tolerant, and immune to electromagnetic fields[3,4]. All-fiber interferometric techniques like Sagnac interferometer[5,6], Mach-Zehnder interferometer[7], Michelson interferometer[8] have been reported for distributed contactless sensing of vibrations driven by dynamic displacements and strains. However, these sensors are bulky, require precise alignments, and require multiple sensing channels. Besides, vibration detection has been demonstrated by measuring the external vibration-induced back-scattered light in conventional fiber using optical time-domain reflectometry (OTDR) techniques that rely on the modulation of the phase (Φ-OTDR)[9,10], polarization state (POTDR)[11], and beam frequency (BOTDR)[12] of the propagating wave in the presence of external perturbation. However, these sensing systems incur diminished sensitivity over long-range distributed sensing and require additional expensive components like pulsed lasers, electro-optic modulators (EOM), and erbium-doped fiber amplifiers (EDFA). Also, FBG based vibration interrogators have been utilized to develop distributed vibration sensors[13]. However, FBG based sensors require additional reference mass that make them bulky and are intolerant towards external temperature fluctuations[14]. Dynamic measurement of instantaneous frequency, amplitude, and phase of the vibration at multiple locations using a single fiber channel has remained challenging and elusive. Simultaneous instantaneous phase monitoring at multiple locations can enlighten the complex structural vibrations associated with unsteady aerodynamic loads and can also be used to infer instantaneous peak vibrations, which are otherwise difficult to be monitored using conventional Fourier transform-based averaging techniques[15].

In this letter, we propose and demonstrate a multipoint vibration sensing technique for detection of instantaneous frequency, amplitude, and phase of vibrations by implementing photonic crystal fiber (PCF) based modal interferometery principle. Such interferometry principle involves fabricating fiber configurations with specialty waveguides that enables excitation and recombination of waveguide modes resulting in stable interference spectra over the source spectrum. In our approach, commercially available solid-core photonic crystal fiber (SCPCF) based identical interferometers are concatenated along single fiber channel (fig.1) for detecting mechanical vibrations at multiple locations.

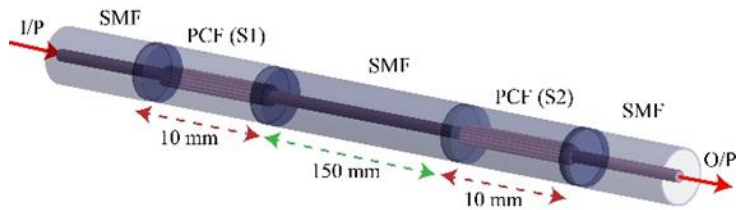

FIG. 1. Schematic representation of integrated sensor probes for vibration sensing. Each PCF section acts a modal interferometer to enable sensing.

The PCF sections act as independent modal interferometers wherein the interference between the PCF's core and higher-order cladding modes occurs. The resultant intensity variation over the source spectrum at each modal interferometer, $I_j$, can be written as[16],

$$I_j = I_{cr} + I_{cl} + 2\sqrt{I_{cr}I_{cl}}\cos\phi \quad (1)$$

where, $I_{cr}$ and $I_{cl}$ are spectral intensity distributions of core and cladding modes of PCF, φ is the phase difference between the two modes given by,

$$\phi = \frac{2\pi L(n_{cr}^{eff} - n_{cl}^{eff})}{\lambda} = \frac{2\pi \Delta n L}{\lambda} \quad (2)$$

where, $L$ is the length of PCF, $\lambda$ is the free space wavelength of light, $n_{cr}^{eff}$ and $n_{cl}^{eff}$ are the effective refractive indices of core and cladding modes of PCF respectively and $\Delta n$ is the difference between the effective modal refractive indices. The maxima of the spectra occur for the condition $\varphi=2N\pi$ or $\Delta nL=N\lambda$ that is dependent on the wavelength of light and length of the interferometer at a constant value of $\Delta n$ and $N$ is an integer. In the presence of an external strain field, the length of the interferometer changes due to field-induced bending, and the magnitude of change in length depends on the amplitude and direction of applied strain. Due to a change in interferometer length, the wavelength values corresponding to the interference maxima condition readjust. This causes shift in the maxima peak positions that results in modulation of $I_j$ over the spectral range. In the presence of alternating strain or vibrational field of specific amplitude and frequency about an interferometer, the bending of the interferometer becomes periodic. As a result, the interference maxima shifts periodically over a range of wavelength values resulting in dynamic modulation of $I_j$. The corresponding sensitivity ($S_j$), that is the net shift of maxima wavelength of $I_j$ with respect to applied measurand field, for each interferometer ($j$) is represented as,

$$S_j = \frac{\partial I_j}{\partial X_j} \quad (3)$$

where, $X$ is the external vibration field that can be given as $X = X_o \sin(\omega t + \theta)$ with $X_o$, $\omega$, and $\theta$ being the amplitude, frequency, and initial phase of vibration, respectively. Under the simultaneous operation of multiple concatenated interferometers in transmission mode, the resultant sensitivity is a summation of the individual sensitivities and can be represented as,

$$S_m = \sum_j \frac{\partial I_j}{\partial X_j} \qquad (4)$$

where, $j = 1, 2, ...$ Thus, the transmitted signal is a superposition of the individual interferometer responses. The resultant real-time sensitivity is influenced by the amplitude, frequency, and relative phase of the vibrations about each interferometer. As proof of concept, we demonstrate multipoint vibration sensing using two identical independently operating modal interferometers integrated along a single fiber length. The sensor probes are kept in contact with externally controlled vibrations and characterized by the wavelength interrogation technique.

To fabricate the sensor probes, two solid-core photonic crystal fiber (SCPCF) sections of length $10 \pm 0.01$ mm are taken with their ends being sharply cleaved using a precision fiber cleaver. The PCF sections are then spliced sequentially along a standard Single-Mode Fiber (SMF) using a programmable fusion splicer at 150 mm apart along the SMF length as shown in fig. 1. Each PCF section acts as a sensor probe to detect the external vibrations. As the interferometers operate independently, the distance between the interferometers can be chosen as per the sensing positions. To characterize the sensor probes, light from a broadband Super-luminescent Light Emitting Diode (SLED) source emitting about a central wavelength of 1552 nm is launched through the in-line interferometers. The resultant interference spectrum is recorded using a spectrum analyzer (OSA).

The spectrum (Fig.2) is first recorded for a single interferometer (S1) (red line) and then for the two interferometers integrated in series (blue line).

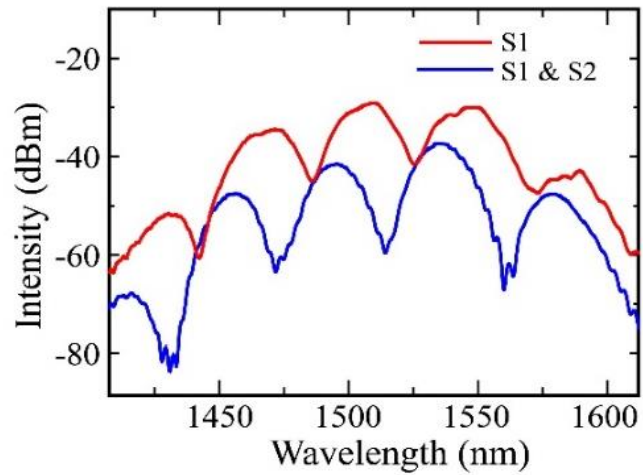

FIG. 2. Characteristic interference spectra of the sensors.

As the interferometers are identical, the spectrum of concatenated interferometers is similar to that of single interferometer except with weak overriding peaks in the former that arises from the intermodulation products in typical sensor array. The negligible loss in the average power is attributed to the splice loss at the splicing points. To interrogate the interferometers with respect to external vibrations over real-time, the interferometers, S1 and S2, are mounted on piezoelectric transducers, PZT (T1 and T2, respectively) as shown in fig.3. The PZTs are powered by a multi-channel piezoelectric controller. To generate vibrations of varied frequencies and amplitudes, the piezo controller is driven by a function generator to control the vibration parameters of the PZTs. The sensor probes are kept straight to avoid strain or bend-induced losses along the interferometers' fiber length. For detecting the real-time signal, the transmitted spectra are analyzed with an FBG based wavelength interrogator, and the output is recorded on a computer. The periodic shift in maxima wavelength of transmission spectra, in presence of external vibrations, is tracked for recording the real-time response of sensor probes.

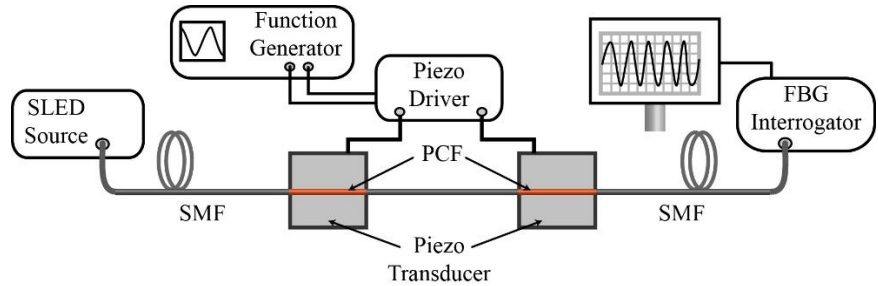

FIG. 3. Schematic diagram of the experimental set-up.

The parameters of vibrations that is amplitude, frequency, and initial phase are varied about each interferometer and the signals are recorded for analysis. Fig. 4 shows the results of the characterization of the sensing system. Firstly, in-phase vibrations of 50 Hz are applied about both interferometers and only the amplitude of vibrations at S2 is varied (see Fig. 4 (a)). From the calibrated curve it is observed that the sensitivity increases linearly with the rise in applied vibrational amplitude that enables detection of minute changes in vibrational amplitude over broad range. Similar observations are made when vibrations of varying amplitudes are applied only about S1 keeping the amplitude about S2 fixed. Subsequently, the values of amplitude and initial phase of vibrations about both interferometers are fixed at 1.5 V and 0° respectively. The frequency of vibrations at S1 is fixed at 5 Hz while the vibration frequency is varied about S2 (see, fig. 4 (b), (c), and (d)). It is observed from the time-series signal that the sensitivity at S1 is modulated by the sensitivity at S2 and the resultant periodic shift in the interference spectra is a superposition of the individual signals, as predicted by theoretical analysis. The complex signal is decomposed into individual components using Wavelet transform (WT) signal processing tool. Using WT methods, individual frequencies are extracted, and their instantaneous phase is evaluated (see fig. 4 (e), (f), and (g)). The fast fourier transform (FFT) of the complex signals are computed and depicted in fig. 4(h). The FFT shows sharp peaks at each frequency with a high S/N (Signal to Noise) ratio.

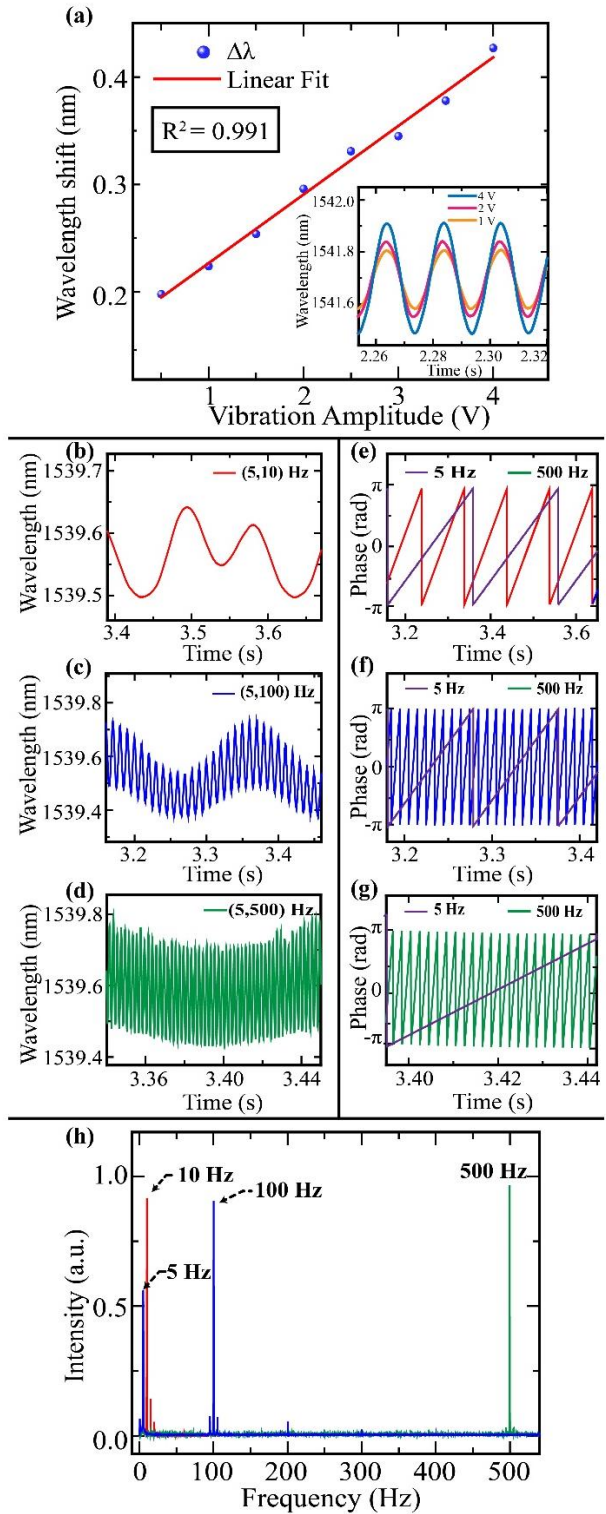

Fig.4: (a) Characteristic plot for wavelength shift of resultant signal versus vibration amplitude. (inset) Real time resultant signal for varying only the amplitude of vibrations applied about S2 while keeping all other parameters identical about S1 and S2. (b)-(d) Real time resultant signal for varying only the frequency of vibrations applied about S2 while applying vibrations of 5 Hz about

S1 and keeping all other parameters identical about S1 and S2. (e)-(g) Instantaneous phase of individual signals about S1 and S2 for the signals shown in (b)-(d). (h) FFT of the signals represented in (b)–(d) that show sharp peaks at the applied vibration frequencies.

Similar observations are made when the frequency of vibrations is kept fixed about S2 and is varied at S1. Thus, the sensing system enables monitoring the instantaneous frequency, amplitude, and phase of each vibration independently about multiple points. Further, we investigated the response of the proposed sensing system for applying vibrations of constant frequency and constant amplitude about S1 and S2 but with different initial phases about S1. It is observed that the resultant real-time sensitivity decreases with an increase in the relative phase difference between the vibrations about S1 and S2 as shown in fig.5. It is maximum when the relative initial phase is 0° and minimum for that of 180°. The cause is attributed to the constructive and destructive modulation of the individual signals at 0° and 180° relative phase between the vibration signals, respectively.

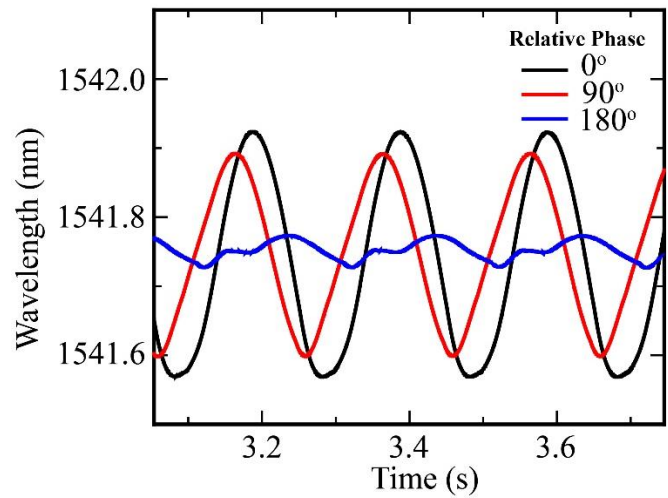

Fig.5: Real time resultant signal for applying vibrations of 10 Hz and fixed amplitude (1.5 V) but with initial relative phase difference about S1 and S2. The relative phase difference between the vibrations is provided in the figure.

In case of operating both the sensors simultaneously, the sensing system's real-time response depends on the sequence of the applied vibration frequencies. Figure 6 illustrates the time series outputs for first applying vibrations of 5 Hz at S1 and 10 Hz at S2 (orange curve) and subsequently reversing the order of vibrations (blue curve). The signals are distinct, and hence the sequence can be identified using suitable pattern recognition techniques.

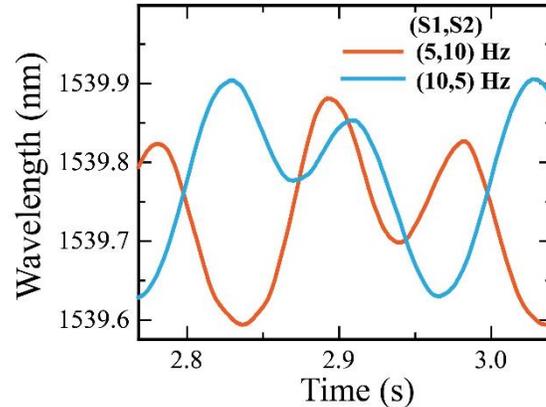

Fig.6 Real time response for operating S1 at 5 Hz and S2 at 10 Hz (orange curve) and for operating S1 at 10 Hz and S2 at 5 Hz (blue curve).

In conclusion, we proposed and demonstrated a compact multipoint vibration sensing technique enabled by concatenated modal interferometers for simultaneous monitoring of instantaneous frequency, amplitude, and phase of vibration along a single fiber channel. Such a system can be further developed with more than two interferometers in series. The sequence of vibrations can also be identified for the localization of source of vibrations. The interferometers are identical and operate independently to generate a resultant output possessing an signal of perturbing field at different locations. Besides, the interferometers' size can be reconfigured while keeping them identical, and the distance between their positions can be altered as per requirement. The modal interferometry technique for distributed sensing offers a platform for designing robust and sensitive single-channel vibration sensing over long distances as an alternative to the existing expensive commercial techniques for industrial applications and structural surveillance. The proposed system can be tuned as per requirement and can be easily extended for more than two interferometers.

Data Availability: The data that support the findings of this study are available from the corresponding author upon reasonable request.

RJ acknowledges the support of SERB-STAR fellowship from Govt. of India.